# A possible superconducting gap signature with filling temperature around 40 K in hexagonal iron telluride islands

Guanyang He,[1,2] Yuxuan Lei,[1,3,5] Tianheng Wei,[1,3] Yanzhao Liu,[4] and Jian Wang[1,3,6,7*]

[1]*International Center for Quantum Materials, School of Physics, Peking University, Beijing 100871, China*
[2]*ShanghaiTech Laboratory for Topological Physics, ShanghaiTech University, 201210 Shanghai, China*
[3]*Beijing Key Laboratory of Quantum Devices, Peking University, Beijing 100871, China*
[4]*Quantum Science Center of Guangdong–Hong Kong–Macao Greater Bay Area (Guangdong), Shenzhen 518045, China*
[5]*Beijing Academy of Quantum Information Sciences, Beijing 100193, China*
[6]*Collaborative Innovation Center of Quantum Matter, Beijing 100871, China*
[7]*Hefei National Laboratory, Hefei 230088, China*

*Corresponding author.
Jian Wang (jianwangphysics@pku.edu.cn)

**Superconductivity in the iron-chalcogenide series FeSe-Fe(Te, Se)-FeTe has been restricted to the near neighbor of iron selenide (FeSe), with a general consensus that iron telluride (FeTe) is not superconducting. In this study, we report the method to grow FeTe islands with atomically flat surface and hexagonal lattice on $SrTiO_3$ (001) substrates, in which a gap structure with a gap-filling temperature close to 40 K is detected by scanning tunneling spectroscopy. Such signature is examined under various conditions and reminiscent of a superconducting gap structure. This work might offer a potential platform to explore new superconductors at ambient pressure.**

Superconductivity in the iron-chalcogenide family (FeSe and Fe(Te,Se)) was reported over a decade ago.[1,2] It attracted extensive research interest due to the astonishing high-temperature (high-$T_c$) superconductivity observed in monolayer FeSe grown on $SrTiO_3$ (STO) substrates,[3,4] and the topological phase transition triggered by Te substitution in Fe(Te,Se)[5,6]. In contrast, at the opposite end of the iron-chalcogenide series, FeTe was long recognized as a non-superconducting antiferromagnetic metal[7]. After years of investigations on it, superconductivity has been found at the interfaces between FeTe and topological insulators.[8-11] Most recently, two-dimensional superconductivity is observed in single-crystalline FeTe films grown on CdTe (001) substrates by suppressing the monoclinic distortion at low temperature in the tetragonal lattice.[12] Although the epitaxially-grown tetragonal FeTe films on CdTe (001) substrates and the interfaces of tetragonal FeTe and topological insulators are superconducting, the superconductivity in the hexagonal phase of FeTe has not been experimentally reported.

In this work, we report the FeTe islands with hexagonal lattice structure epitaxially grown on STO (001) substrates. Notably, the low-temperature scanning tunneling microscopy/spectroscopy (STM/S) study reveals a gap structure with coherence peak features on the FeTe islands. By increasing the temperature from 4.2 K to 45 K, the gap feature diminishes gradually, consistent with the characteristics of superconductivity. The gap-filling temperature is as high as around 40 K. Such spectral gap is reproducible in multiple hexagonal FeTe islands and with different STM tips. Also, a hole-like band structure is revealed by quasiparticle interference (QPI) on such islands, suggesting hole-type conduction of the hexagonal FeTe islands on STO.

Our experiments are conducted in an MBE-STM combined system (Scienta Omicron, Inc.) of an ultrahigh vacuum of $1\times10^{-10}$ mbar, where FeTe is epitaxially grown on Nb-doped STO (001) (wt 0.7%) substrates. STO is first thermally annealed in vacuum at 1050 °C for 40 minutes to obtain an atomically flat $TiO_2$-terminated surface.

FeTe is grown by co-evaporating high-purity Fe (99.994%) from an electron beam evaporator, and Te (99.999%) from standard Knudsen cells (K-cells). The acceleration voltage and emission current of the electron beam evaporator are 800 V and 30 mA, respectively. By adjusting the growth conditions, we obtain FeTe films with three phases of tetragonal, hexagonal and honeycomb lattices (See details in the Supplementary data). For a growth condition of high Te flux (K-cell temperature of Te is $T_{Te}$ = 310 °C) and high substrate temperature ($T_{sub}$ = 350 °C), atomically flat FeTe islands among the small clusters are obtained after post-annealing for ten minutes at 280 °C, with a typical STM image of an island shown in Fig. 1(a). The STS data is measured at 4.3 K by a polycrystalline Pt/Ir tip and standard lock-in technique. The modulation voltage on the tip is 1 mV at 1.769 kHz. The set-up conditions are V = 40 mV, I = 2.5 nA for tunneling dI/dV measurements, and V = 1 V, I = 0.2 nA for topographic images unless specified otherwise.

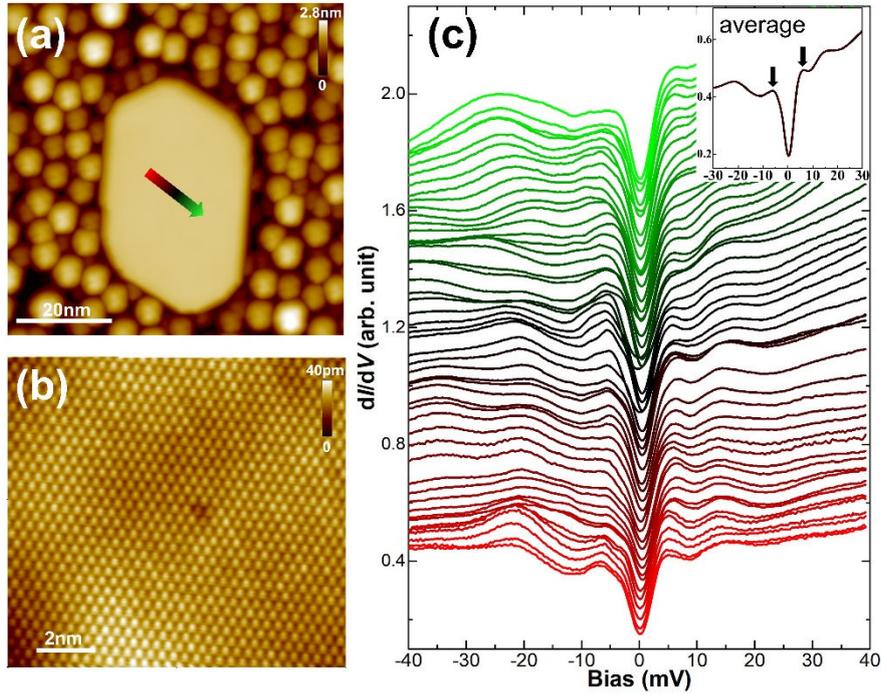

**Fig. 1.** (a) STM topography of an FeTe island among the small clusters. (b) The hexagonal lattice measured on top of the FeTe island. (c) A series of raw spectra measured along the arrow in (a) with color correspondences, giving an averaged spectrum in the inset. Spectra are offset vertically for clarity.

The atomic image of the FeTe island shows hexagonal lattice with the in-plane lattice constant $a_0$ ~ 0.38 nm as shown in Fig. 1(b), exhibiting an incommensurate epitaxial growth on the tetragonal STO substrates (STO lattice constant ~ 0.39 nm). [13] The FeTe island thickness is around four unit cells (see Fig. S2 in Supplementary data), considering the out-of-plane lattice constant of hexagonal FeTe near 0.536 nm.[14] The particle-hole symmetric gap (i.e., the positions of the two STS peaks show equal distance to the Fermi level) in the spectra uniformly appear throughout the island, as shown in Fig. 1(c). The averaged spectrum is exhibited in the inset of Fig. 1(c), where coherence peak features at ±6 mV are clearly observed as indicated by black arrows, reminiscent of a superconducting gap structure.

To investigate the temperature dependence of the observed gap structure, dI/dV spectra at varying temperatures from 4.2 K to 45 K on a hexagonal FeTe island are measured. The tunneling spectra (hollow dots) normalized by their background curves (see details in Supplementary data) are shown in Fig. 2. The coherence peak features are progressively weakened and smeared with increasing temperature, and the gap feature almost disappears around 45

K, showing a high gap-filling temperature around 40 K. Such temperature evolution behavior is further confirmed in another hexagonal FeTe island (see Fig. S4), indicating the possible characteristic of superconductivity.

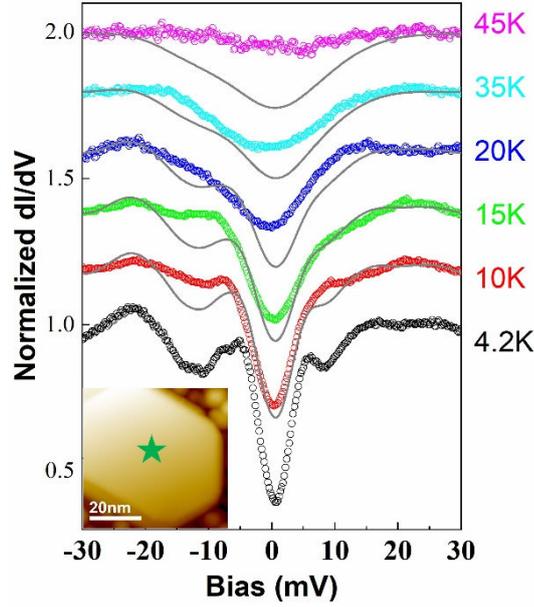

**Fig. 2.** The hollow dots with different colors are the normalized experimental spectra at different temperatures, and the grey curve at each temperature higher than 4.2 K is the convolution of 4.2 K spectrum and the Fermi–Dirac distribution function. The spectra at varying temperatures are measured at the green star position on a hexagonal FeTe island shown in the inset. The hollow dots spectra with the grey curves are offset vertically for clarity.

In order to examine the thermal smearing effect in experiments, the STS spectrum measured at 4.2 K is convoluted with the Fermi–Dirac distribution function, to simulate the thermal broadening spectra at each temperature higher than 4.2 K, [15] as shown by the grey curves in Fig. 2. The experimentally obtained spectra are apparently different from the thermal broadening spectra at higher temperatures, excluding the trivial origin for the observed gaps. The particle-hole symmetric STS gap with the coherent peaks can be detected in multiple hexagonal FeTe islands (see Fig. 3) and further revealed by using the superconducting STM tip (see Fig. S5 in Supplementary data). All these results collectively suggest the observed gap structure might be a signature of superconductivity.

Quantum well states or Coulomb blockade sometimes could cause gap-like dI/dV spectral structures. The positions of spectral peaks of quantum well states are periodic in the bias voltage, and not necessarily symmetric about the Fermi level, [16] which is different from our observations of a pair of symmetric spectral peaks about the Fermi level and the absence of periodic peak oscillations at bias voltages. For Coulomb blockade effect scenario, external electrons should tunnel into the island one at a time under bias voltages, leading to equidistant conductance peaks in the dI/dV spectra, which is also not necessarily particle-hole symmetric and different from our observations. Moreover, the size of Coulomb gaps (separation between conductance peaks) typically increases linearly with the reciprocal of island area. [17] In our work, the gap sizes between two symmetric coherent peaks around the Fermi level in multiple FeTe islands are measured in multiple FeTe islands (partially shown in Fig. 3). No correlation is found between the gap size and the reciprocal of island area (Fig. S3 in Supplementary data), which further excludes the possibility of Coulomb blockade. The spectral gap originating from an antiferromagnetic (AFM) ground state might be another factor to consider. However, the spectral gap due to an AFM order has a clearly asymmetric structure about the Fermi level, [18] not analogous to our observation in FeTe islands.

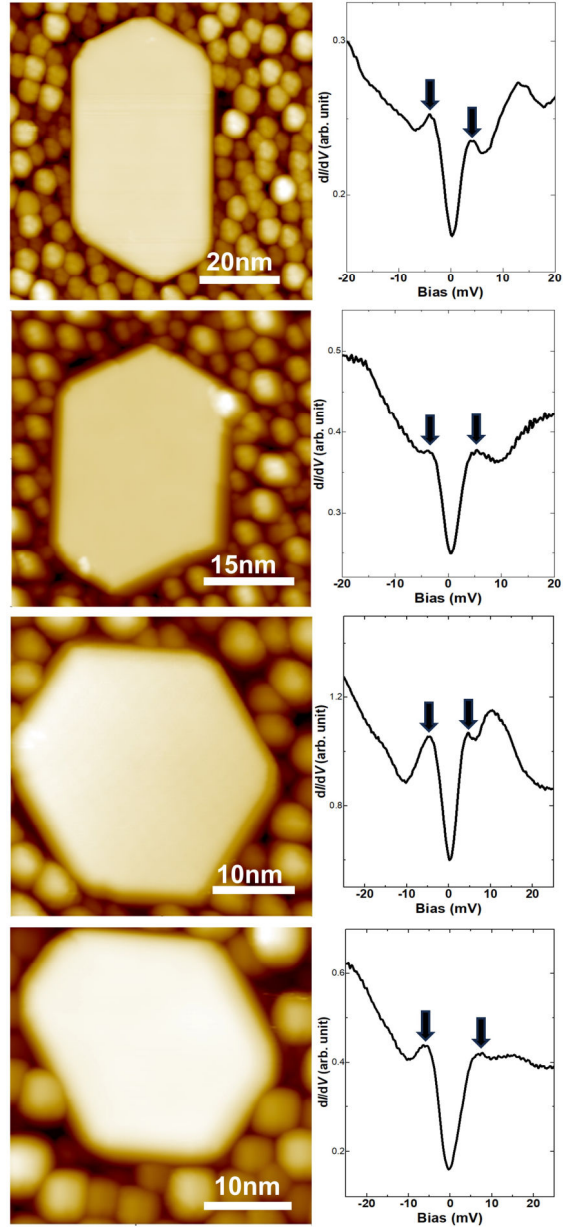

**Fig. 3.** Gap structures measured on multiple FeTe islands. Each spectrum is the average of 3-5 spectra measured at the center of the FeTe island shown to the left of the spectrum. The black arrows indicate the positions of coherence peak signatures, based on which the gap size is estimated.

The electronic band structure of hexagonal FeTe islands is revealed by QPI. We measure the dI/dV mapping on an FeTe island as shown in Fig. 4(a), with the momentum-space structure shown by its Fourier transform in Fig. 4(b). A ring-like structure due to the scattering intensity from quasiparticles on a constant-energy circle can be roughly seen in the momentum space. After suppressing irrelevant scattering intensities of small scattering vectors **q**, the radius of constant-energy circle is found to shrink with increasing energy in Fig. 4(c). We average the intensity of same radius |**q**| and plot the scattering intensity in a |**q**|-E plane in Fig. 4(d). A hole-like band structure is manifested by the parabolic fitting (black dashed line) to the intensity. It is consistent with the calculated hole band structure of hexagonal FeTe in a theoretical work, [19] which predicts a possible magnetically-mediated superconductivity in hexagonal FeTe.

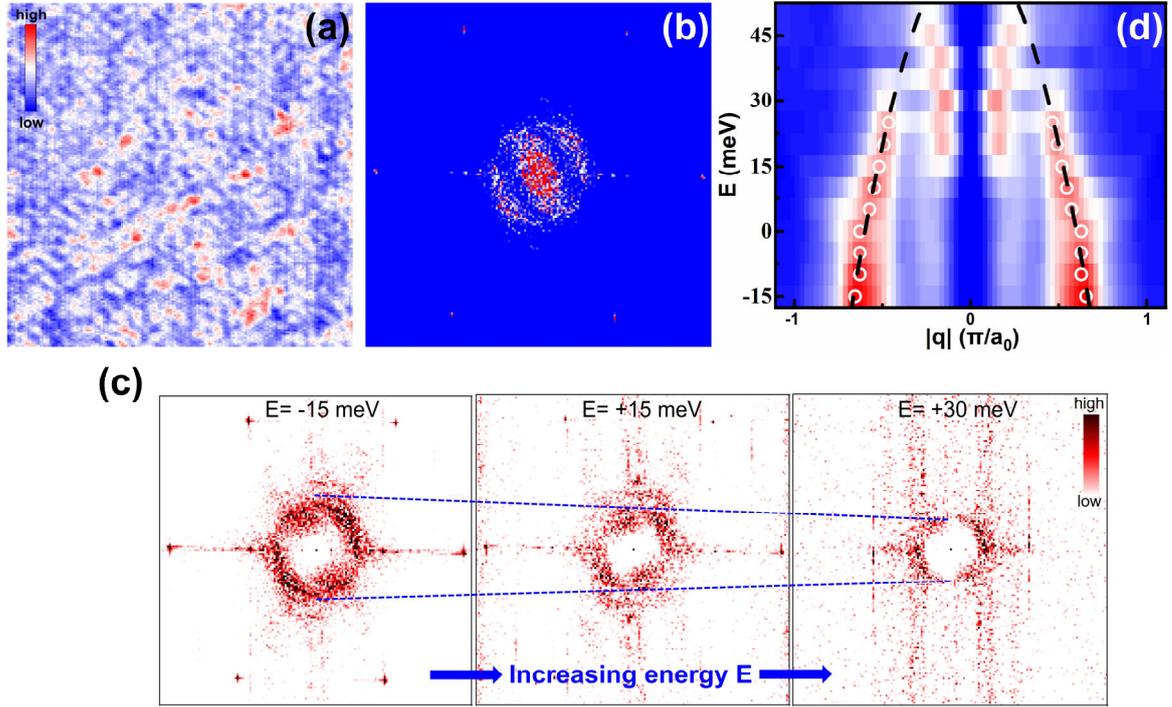

**Fig. 4.** (a) dI/dV mapping of a 26.5×26.5 nm$^2$ area on a hexagonal FeTe island at energy E = -15meV. (b) Fourier transform of (a) showing the scattering intensity in the momentum space. (c) Scattering intensity from quasiparticles on a constant-energy circle in the momentum space. The circle radius varies with energy. (d) Scattering intensity plotted in the |**q**|-E plane, showing the band structure. The hollow circles and a dashed line indicate the high intensity positions and a parabolic band fitting, respectively.

In summary, particle-hole symmetric gaps with coherence peak features are detected by STM/S measurements on the hexagonal FeTe islands grown on STO (001) substrates with the thickness of around four unit cells. The gap structure is suppressed with the increasing temperature, and the quantum well effect, Coulomb blockade and AFM origins for the spectral structure can be ruled out to a certain extent. The particle-hole symmetric gap structure with coherence peaks is repeatably observed in multiple FeTe islands via non-superconducting or superconducting STM tips. Our observations and analyses consistently suggest the observed gap structure on hexagonal FeTe islands might be a signature of superconductivity, which still awaits in-depth evidences such as transport results, diamagnetic response, and vortex states under external magnetic fields. These experiments are absent in this work due to the lack of magnetic field capability in our STM equipment, and transport measurements are impracticable for our present samples due to the locality of FeTe islands. The discovery of the particle-hole symmetric gap structure of the hole-type conducting hexagonal FeTe islands on tetragonal STO surface with the filling temperature around 40 K might shed new light on investigating the interfacial or size-confined quantum phenomena such as superconductivity.

References:

1. Xianhui Chen, Pengcheng Dai, Donglai Feng, Tao Xiang, and Fu-Chun Zhang, Iron-based high transition temperature superconductors. *National Science Review* **2014,** 1, 371.

2. Fong-Chi Hsu, Jiu-Yong Luo, Kuo-Wei Yeh, Ta-Kun Chen, Tzu-Wen Huang, Phillip M. Wu, Yong-Chi Lee, Yi-Lin Huang, Yan-Yi Chu, Der-Chung Yan, and Maw-Kuen Wu, Superconductivity in the PbO-type structure α-FeSe. *Proceedings of the National Academy of Sciences* **2008,** 105, 14262.

3. Qing-Yan Wang, Zhi Li, Wen-Hao Zhang, Zuo-Cheng Zhang, Jin-Song Zhang, Wei Li, Hao Ding, Yun-Bo Ou, Peng Deng, Kai Chang, Jing Wen, Can-Li Song, Ke He, Jin-Feng Jia, Shuai-Hua Ji, Ya-Yu Wang, Li-Li Wang, Xi Chen, Xu-Cun Ma, and Qi-Kun Xue, Interface-induced high-temperature superconductivity in single unit-cell FeSe films on SrTiO3. *Chinese Physics Letters* **2012,** 29, 037402.

4. Wen-Hao Zhang, Yi Sun, Jin-Song Zhang, Fang-Sen Li, Ming-Hua Guo, Yan-Fei Zhao, Hui-Min Zhang, Jun-Ping Peng, Ying Xing, Hui-Chao Wang, Takeshi Fujita, Akihiko Hirata, Zhi Li, Hao Ding, Chen-Jia Tang, Meng Wang, Qing-Yan Wang, Ke He, Shuai-Hua Ji, Xi Chen, Jun-Feng Wang, Zheng-Cai Xia, Liang Li, Ya-Yu Wang, Jian Wang, Li-Li Wang, Ming-Wei Chen, Qi-Kun Xue, and Xu-Cun Ma, Direct Observation of High-Temperature Superconductivity in One-Unit-Cell FeSe Films. *Chinese Physics Letters* **2014,** 31, 017401.

5. Xun Shi, Zhi-Qing Han, Pierre Richard, Xian-Xin Wu, Xi-Liang Peng, Tian Qian, Shan-Cai Wang, Jiang-Ping Hu, Yu-Jie Sun, and Hong Ding, $FeTe_{1-x}Se_x$ monolayer films: towards the realization of high-temperature connate topological superconductivity. *Science Bulletin* **2017,** 62, 503.

6. Chaofei Liu and Jian Wang, Heterostructural one-unit-cell FeSe/SrTiO3: from high-temperature superconductivity to topological states. *2D Materials* **2020,** 7, 022006.

7. Zhimo Zhang, Min Cai, Rui Li, Fanqi Meng, Qinghua Zhang, Lin Gu, Zijin Ye, Gang Xu, Ying-Shuang Fu, and Wenhao Zhang, Controllable synthesis and electronic structure characterization of multiple phases of iron telluride thin films. *Physical Review Materials* **2020,** 4, 125003.

8. S. Manna, A. Kamlapure, L. Cornils, T. Hänke, E. M. J. Hedegaard, M. Bremholm, B. B. Iversen, Ph Hofmann, J. Wiebe, and R. Wiesendanger, Interfacial superconductivity in a bi-collinear antiferromagnetically ordered FeTe monolayer on a topological insulator. *Nature communications* **2017,** 8, 14074.

9. Hemian Yi, Yi-Fan Zhao, Ying-Ting Chan, Jiaqi Cai, Ruobing Mei, Xianxin Wu, Zi-Jie Yan, Ling-Jie Zhou, Ruoxi Zhang, Zihao Wang, Stephen Paolini, Run Xiao, Ke Wang, Anthony R. Richardella, John Singleton, Laurel E. Winter, Thomas Prokscha, Zaher Salman, Andreas Suter, Purnima P. Balakrishnan, Alexander J. Grutter, Moses H. W. Chan, Nitin Samarth, Xiaodong Xu, Weida Wu, Chao-Xing Liu, and Cui-Zu Chang, Interface-induced superconductivity in magnetic topological insulators. *Science* **2024,** 383, 634.

10. Qing Lin He, Hongchao Liu, Mingquan He, Ying Hoi Lai, Hongtao He, Gan Wang, Kam Tuen Law, Rolf Lortz, Jiannong Wang, and Iam Keong Sou, Two-dimensional superconductivity at the interface of a Bi2Te3/FeTe heterostructure. *Nature communications* **2014,** 5, 4247.

11. Xiong Yao, Hee Taek Yi, Deepti Jain, Xiaoyu Yuan, and Seongshik Oh, Mystery of superconductivity in FeTe films and the role of neighboring layers. *APL Materials* **2025,** 13, 011116.

| 12 | Yuki Sato, Soma Nagahama, Shunsuke Kitou, Hajime Sagayama, Ilya Belopolski, Ryutaro Yoshimi, Minoru Kawamura, Atsushi Tsukazaki, Naoya Kanazawa, Takuya Nomoto, Ryotaro Arita, Taka-hisa Arima, Masashi Kawasaki, and Yoshinori Tokura, Superconductivity and suppressed monoclinic distortion in FeTe films enabled by higher-order epitaxy. *Nature communications* **2025,** 16, 10913. |
|---|---|
| 13 | Guanyang He, Yu Li, Yuxuan Lei, Xingyue Wang, Minghu Pan, and Jian Wang, Impacts of substrate conditions and post-annealing on monolayer iron-based superconductors. *Physical Review Materials* **2024,** 8, 014802. |
| 14 | Lixing Kang, Chen Ye, Xiaoxu Zhao, Xieyu Zhou, Junxiong Hu, Qiao Li, Dan Liu, Chandreyee Manas Das, Jiefu Yang, Dianyi Hu, Jieqiong Chen, Xun Cao, Yong Zhang, Manzhang Xu, Jun Di, Dan Tian, Pin Song, Govindan Kutty, Qingsheng Zeng, Qundong Fu, Ya Deng, Jiadong Zhou, Ariando Ariando, Feng Miao, Guo Hong, Yizhong Huang, Stephen J. Pennycook, Ken-Tye Yong, Wei Ji, Xiao Renshaw Wang, and Zheng Liu, Phase-controllable growth of ultrathin 2D magnetic FeTe crystals. *Nature communications* **2020,** 11, 3729. |
| 15 | Cheng Chen, Kun Jiang, Yi Zhang, Chaofei Liu, Yi Liu, Ziqiang Wang, and Jian Wang, Atomic line defects and zero-energy end states in monolayer Fe (Te, Se) high-temperature superconductors. *Nature Physics* **2020,** 16, 536. |
| 16 | Chaofei Liu, Chunxiang Zhao, Shan Zhong, Cheng Chen, Zhenyu Zhang, Yu Jia, and Jian Wang, Equally Spaced Quantum States in van der Waals Epitaxy-Grown Nanoislands. *Nano Letters* **2021,** 21, 9285. |
| 17 | Guanyang He, Yu Li, Yuxuan Lei, Andreas Kreisel, Brian M. Andersen, and Jian Wang, Lateral Quantum Confinement Effect on High-$T_C$ Superconducting FeSe Monolayer. *Nano Letters* **2024,** 24, 7654. |
| 18 | Xiaodong Zhou, Peng Cai, Aifeng Wang, Wei Ruan, Cun Ye, Xianhui Chen, Yizhuang You, Zheng-Yu Weng, and Yayu Wang, Evolution from Unconventional Spin Density Wave to Superconductivity and a Pseudogaplike Phase in $NaFe_{1-x}Co_xAs$. *Physical Review Letters* **2012,** 109, 037002. |
| 19 | David S. Parker, Strong 3D and 1D magnetism in hexagonal Fe-chalcogenides FeS and FeSe vs. weak magnetism in hexagonal FeTe. *Scientific Reports* **2017,** 7, 3388. |